# The Role of Microtubule Movement in

# Bidirectional Organelle Transport

Igor M. Kulić[1], André E.X. Brown[2], Hwajin Kim[3], Comert Kural[4], Benjamin Blehm[4], Paul R. Selvin[4], Philip C. Nelson[2] and Vladimir I. Gelfand[3]

1 School of Engineering and Applied Sciences,  29 Oxford St., Pierce Hall 409, Harvard University, Cambridge, MA 02138 Tel: (1) 617-496-9033, Fax: (1) 617-495-9837 email: kulic@seas.harvard.edu (corresponding author)

2 Department of Physics and Astronomy and Nano/Bio Interface Center, University of Pennsylvania, Philadelphia, PA 19104

3 Department of Cell and Molecular Biology, Feinberg School of Medicine, Northwestern University, Chicago, IL 60611

4  Physics Department and  Biophysics Center, University of Illinois at Urbana-Champaign, Urbana, IL 61801

## *Abstract*

We study the role of microtubule movement in bidirectional organelle transport in *Drosophila* S2 cells and show that EGFP-tagged peroxisomes in cells serve as sensitive probes of motor induced, noisy cytoskeletal motions. Multiple peroxisomes move in unison over large time-windows and show correlations with microtubule tip positions, indicating rapid microtubule fluctuations in the longitudinal direction. We report the first high resolution measurement of longitudinal microtubule fluctuations performed by tracing such pairs of co-moving peroxisomes. The resulting picture shows that motor-dependent longitudinal microtubule oscillations contribute significantly to cargo movement along microtubules. Thus, contrary to the conventional view, organelle transport cannot be described solely in terms of cargo movement along stationary microtubule tracks, but instead includes a strong contribution from the movement of the tracks.

## Abbreviation List

MSD (mean square displacement), mCherry (monomeric Cherry fluorescent protein), TIRF (total internal reflection fluorescence)

## *Introduction*

Molecular motor mediated transport along microtubules is an extensively studied phenomenon *in vitro*[1][3] . Despite significant advances *in vitro*, understanding how intracellular transport works *in vivo* still remains one of the big challenges in cell biology. The questions of how cellular cargos find their way through the cytoplasm and get targeted to their temporary or final destinations lie at the heart of the problem. One of the major puzzles in this context is the so-called bidirectional organelle transport. The majority of cargos in the cell move in a bidirectional and often remarkably symmetric manner [4][5]. Despite the known kinetic and dynamic asymmetry of the underlying plus- and minus-end directed microtubule motors, the vesicles seem to move with the same rates and run length distributions, and exhibit identical stalling forces, in each direction [7].  Furthermore, inhibition of transport in one direction typically results in the inhibition of movement in the opposite direction as well [4]-[5].

One straightforward explanation of bidirectional organelle transport rests upon the hypothesis [4][5] that a dedicated molecular mechanism couples opposite polarity microtubule motors in vivo.  While this is possible, we suggest an alternative, perhaps complementary hypothesis, to the motor coupling hypothesis. Our hypothesis rests on the plausible assumption that a cargo vesicle has multiple motors residing on it, and these couple to several microtubules at a time. The conflicting strains cause them to slide, bend and buckle, causing effective aperiodic limited amplitude fluctuations. These fluctuations modify the motion of vesicles in an additive manner, contributing to the phenomenology of bidirectional organelle transport. We describe unusual observations coming from high resolution traces of single peroxisomes, in particular, the unusual mean square displacement behavior and large velocity cross-correlation observed between peroxisomes.  We then present results from simultaneous two-color imaging of peroxisomes and microtubules. It is shown that in many cases a strong correlation between the peroxisomes and microtubules can be established. This finally leads us to a proposed model that integrates the dynamic interplay of vesicle motion, molecular motor action, and microtubule motility.

## *Results*

### Behavior of single and multiple peroxisomes in vivo

In a previous study of GFP labeled peroxisome motion in *Drosophila* S2 cells, Kural et al.[12] demonstrated a velocity distribution exhibiting a significant contribution of very large velocities (>10 µm/s) over larger time intervals (several tens of milliseconds).  We have reanalyzed some of these extreme velocity events. Often such fast movements were preceded or followed by rapid movements in the opposite direction (in 32 out 36 or 89% of trajectories, lasting for at least 10 s) with sub-second time intervals between direction switching events-- a characteristic signature of bidirectional transport.

While non-steady relaxing velocities were characteristic of 89% of 36 motile particle trajectories , single exponential velocity decays over intervals larger than 100 ms as shown in Fig. 1a (inset, blue curve) were rare (in 9 out of 36 or 25% of trajectories, cf. supporting materials).  Instead, a more complex behavior, including the superposition of multiple relaxation times and direction switching events, was predominant (Fig 1a,b). These velocity relaxation events indicated the presence of an elastic component in the system and suggested that bent and buckled microtubules could influence peroxisome transport. This hypothesis was further strengthened by the observation of several vesicles moving in concert (Fig. 1b) with strong velocity cross-correlation on timescales larger than 30 ms (Fig. 1b inset).

Because of the limited number of peroxisomes observed in close proximity to other peroxisomes (14 observations of peroxisomes closer than 5µm), the observation of co-moving peroxisome pairs was rare (3 observations). This is consistent with the expectation that the large number of microtubules per process ($N_{MT}$~5-10, determined by counting the microtubules converging and entering the processes) reduces the probability of two peroxisomes to be found on the same microtubule (p= 1/( $N_{MT}$ -1)~10-25% ). However those peroxisomes moving in concert stayed in the highly correlated state for longer than our observation time of 20 seconds (10000 frames). Despite the large (>90%) velocity correlation of peroxisome pair-speeds (cf. Fig 1b inset), their relative distance was not strictly constant, and in fact, was slowly changing in time (by 220 nm over 20 sec).

   A systematic analysis of peroxisomes in thin processes of S2 cells showed two different types of moving behaviors: 1) A population of relatively immobile particles (25 of 61 or ~40 % of the total population), moving less than 100 nm during a 5 second-interval, whose trajectories did not exhibit clear alignment with the process/microtubule axis; 2) A rapidly moving population (36 of 61 or ~60%) whose trajectories were parallel to the process axis (cf. methods).

We focused on this motile population (2) showing large displacements. We analyzed the mean square displacement $MSD(\Delta t) = \left\langle (x(t+\Delta t) - x(t))^2 \right\rangle$ of single peroxisome traces as a function of the time lag $\Delta t$ (i.e. the difference between two time points).  If the peroxisomes moved in a directed constant-velocity fashion we would expect $MSD(\Delta t) \propto \Delta t^2$, i.e. a purely "ballistic" scaling behavior whereas if they moved in purely random self-uncorrelated fashion we would expect a "diffusive" scaling $MSD(\Delta t) \propto \Delta t$. If on short timescales the motion was strictly directed and driven by either kinesin or dynein at any given time, but on longer timescales a random switching between them occurred, then we would expect on short times $MSD(\Delta t) \propto \Delta t^{\alpha}$ with an exponent $\alpha = 2$, and for longer time-lags $\Delta t$ a cross-over to diffusive scaling behavior $MSD(\Delta t) \propto \Delta t^{\beta}$ with $\beta = 1$ (cf. the supporting material).  In this case the cross-over time for the switch from constant velocity to diffusive behavior would be interpreted as the typical switching time between kinesin and dynein.

   Contrary to this naïve expectation we observed an unexpected distribution of mean square displacement exponents that clearly deviated from β = 1 and α = 2 (Fig 1c).  The trajectories of 36 particles from 9 different cells were analyzed and their MSD as function of the time lag was calculated. Generically the majority of trajectories (N=32) showed two clearly distinct scaling regimes (Fig 1c inset). At short time-lags $\Delta t < 30\ ms$, the peroxisomes demonstrate sub-diffusive behavior $MSD(\Delta t) \propto \Delta t^{\alpha}$ ($\alpha < 1$) with a scaling exponent $\alpha = 0.59 \pm 0.28$ exhibiting a single broad peak at ~0.5 (Fig. 1d). The relatively small squared displacements (typically ~100nm$^2$) and the lack of correlation between one peroxisome and another (see Fig. 1B), moving in concert on this short timescale (as opposed to their high correlation on long timescales) indicate that local environment effects and thermal fluctuations dominate the peroxisome motion at very short timescales rather than an active driving force (i.e. microtubule motors). At longer timescales $30\ \text{ms} < \Delta t < 3\ \text{s}$, single peroxisomes exhibit enhanced diffusion $MSD(\Delta t) \propto \Delta t^{\beta}$ ($1 < \beta < 2$) with a bimodal distribution of scaling exponents with two local maxima close to $\beta \approx 1.5$ and 2.0 and an overall mean 1.62 (standard deviation 0.29). This indicated that a certain fraction (~30%) of vesicles was indeed moving with a constant velocity (~2.0 exponent), consistent with a simple model of a cargo hauled by motors on a spatially immobile microtubule. However, the majority of traces (including vesicles moving in concert, Fig 1b,c) showed a sub-ballistic but hyper-diffusive dynamics ($1 < \beta < 2$) with an exponent close to 1.5 (Fig. 1d), an observation challenging the simple motor-hauling-a-cargo and random motor switching model and indicating the movements of microtubules.

## The microtubule motion

To determine the contribution of microtubule movement in vesicle transport, we simultaneously visualized peroxisomes and microtubules by tagging them with EGFP and mCherry fluorescent proteins, respectively (cf. Methods section, [7]). As shown in Fig 2, microtubules in Cytochalasin D-treated S2 cells form bundles in the processes and a loose meshwork in the cell body and the general microtubule pattern remains relatively constant over long periods of time. At the same time, analysis of time-lapse sequences shows that microtubules display large lateral and longitudinal motions both in the cell body and in the processes. The microtubule bundles in processes are confined within a diameter of only 1-2 μm and therefore lateral movements of microtubules often result in their bending and bucking rather than random excursions seen in the cell body. (Figure 2a,b). Given microtubules' high bending stiffness constant B~$2 \times 10^{-23}$Nm$^2$ [13] , this indicates exerted forces in the piconewton range acting on sliding microtubules in longitudinal directions. By measuring the relative sliding speeds from the relative excess length variation of neighboring microtubules (cf. methods section), we were able to estimate the relative sliding velocities of microtubules in bundles. In several cases shown in Fig 2b, microtubules were found to slide relative to each other at typical microtubule-dependent motor velocities, $0.3 - 1.5$ µm/s over timescales of several seconds. More extreme microtubule rearrangements were also observed, including movement of microtubule loops within the processes, which indicate that strong longitudinal shear forces act on the microtubules, presumably due to the action of molecular motors (Figure 2c).

The sliding of microtubules was often related to the motion of single or multiple peroxisomes (Fig. 3). Fig. 3a shows a peroxisome that moves along a microtubule bundle, dynamically "clamping" two microtubules together. Another remarkable observation was the buckling and bending of whole microtubule bundles in close proximity to peroxisomes (Fig. 3b). During the buckling events, the bundles were split into several sub-bundles and single microtubules that converged together at the position of the vesicle. We also observed peroxisome motions that were highly correlated with the motions of distant microtubule tips over extended time intervals longer than 10 s (4 observations) (Fig. 3c), consistent with the observation of correlated motion of peroxisome pairs (Fig. 1b). Fig 3c (left) shows the kymograph of the peroxisome and microtubule tip position exhibiting a high correlation coefficient (0.92). This provides further evidence that microtubule movement affects cargo motion and vice versa.  In some cases, the correlations between microtubule tip and peroxisome positions persisted for up to 10 minutes and were also observed even after blocking microtubule dynamics with Taxol (data not shown).

Notably, the rapid microtubule motions were not restricted to cell processes. In fact, microtubule fluctuations appeared to be even more pronounced in the cell body where lateral microtubule motion is less confined than in the processes. We found unusual non-random crossover points of several microtubules that indicated dynamic cross-linking of 3 or more microtubules that persisted over several minutes (Fig. 4). Although the curvature and shape of the participating microtubules changed dramatically over time, their crosslinking points remained stable while moving over micron distances. Longitudinal velocities of tip movement for microtubules containing cross-linking points (0.5-1 µm/s over timescales of seconds) were significantly larger than the maximal microtubule polymerization speeds measured by tracing EGFP- tagged-EB1 protein particles (~0.2 µm/s).  This indicates that the microtubule cross-linking points, for which we propose the name "hubs", are not static but instead are very dynamic structures which could possibly be the source of active forces for moving microtubules and peroxisomes. Some of the triple-crossover hubs ( i.e. spots where three microtubules come together), showed changes in the number of participating microtubules where one of the microtubules was released and recaptured tens of seconds later and then remained in the hub for the rest of the recording (Fig 4 b, c). While the microtubules in the hubs were moving longitudinally and laterally with typical motor speeds, the position of the hubs remained relatively constant throughout the recording (featuring displacements of < 1 µm over 60 seconds). This indicates that the colocalization of several microtubules in one hub is not simply the consequence of projection of microtubules onto a single image plane but rather a physical cross-linking point between several microtubules.

    Although the molecular origin of cross-linking in the hubs remains unclear, the dynamic nature of cross-linking suggests involvement of motor proteins. While microtubule associations were in some cases caused by peroxisomes (Fig. 3a,b), more frequently hubs did not colocalize with peroxisome positions (Fig. 4). This however is not inconsistent with our hypothesis that hubs are formed by motor decorated vesicles, since peroxisomes constitute only a small fraction of all the cellular cargos.

## *Discussion*

### Origin of dynamic microtubule features in S2 cells

The most striking observations in peroxisome motility in S2 cell processes were: 1) Sharp changes in velocities: initially high but quickly decaying.  2) Hyper-diffusive movement of peroxisomes with a MSD scaling exponent close to 1.5.  3) Peroxisome pairs that moved in concert over large time intervals. The first observation is consistent with vesicle motion driven by the relaxation of an initially bent elastohydrodynamic element, likely a microtubule or microtubule bundle, given the large released lengths and the rapid relaxations. Specifically, the observed relaxation time of 50-1000 ms was consistent with the elastohydrodynamic relaxation timescales $t \sim \eta L^4/B$ that are expected for a microtubule segment of several microns in length ($\eta \sim 10^{-2}$-$10^{-1}$ Pa s is the cytoplasmic viscosity, L the length of the buckled segment, and $B \sim 2 \ 10^{-23}$ Nm$^2$ the microtubule bending stiffness constant [13], cf. supporting information Figs. S4-S6). More generally if the relaxation involves many modes, a power-law scaling of the MSD instead of a single exponential decay is expected and in fact observed here. The observation of vesicles moving in pairs presents further strong evidence that vesicle motion is not solely caused by their separately attached motors, but is in fact associated with the motion of the underlying microtubule track itself.

Although the results show that microtubule rearrangements affect vesicle motion, they do not immediately address the source of this rearrangement. However, based on anecdotal inferences, e.g. Fig. 2A, 3B, the observation that microtubule buckling and sliding often occur in close proximity to peroxisomes, suggests that the microtubule movements could originate from the presence of motors on the surface of vesicles and the motors that are bound simultaneously to several microtubules in a bundle. Indeed, the microtubule sliding velocities measured from buckling events are in the range of typical motor speeds (0.3-1.0 µm/s). Notably a specific knock-down of kinesin heavy chain leads to a dramatic reduction of microtubule motions, cf. supporting information. This further strengthens our hypothesis of rapid motor induced microtubule motions, in agreement with similar observations in other systems [26]-[31].

The observation of moving microtubules leads us to propose a tentative schematic model as shown in Fig. 5.  The majority of microtubules are sparsely cross-linked with each other and with other more rigid cellular structures (like nucleus and cell cortex).  The distinct microtubule motions seem to be mediated by motors on the surface of vesicles that are moving on microtubules at various positions: multiple motors of opposing polarity on vesicles bridge two microtubules at different positions to buckle or slide microtubules (Fig. 5a,b) or multiple motors on a vesicle crosslink multiple microtubules simultaneously  to make jointing points of microtubules , or "hubs" (Fig. 5c). Since motors bound on the surface of vesicles generate forces between the vesicles and microtubules, vesicle motion causes various longitudinal and lateral strains in the microtubule backbone. This results in significant displacements of the microtubules. While these displacements could be limited by the microtubule attachment if any anchor exists and could also be sterically confined within the cellular processes, microtubule excursions can easily reach the range of hundreds to thousands of nm depending on the length of the microtubules involved, and the number of active motors on the bound cargoes. Sometimes, vesicles transiently couple to the same microtubules and move in pairs while microtubules are moving as observed for some peroxisomes in the S2 cells (Fig. 5d).

## Physical origin of the unusual scaling exponents

What physical picture of intracellular transport do our observations suggest?

The majority of peroxisomes, in particular those moving in concert with other peroxisomes, exhibit hyperdiffusive behavior with MSD($\Delta t$) proportional to $\Delta t^{3/2}$. It is experimentally [13]-[15] and theoretically [16]-[20] well established that solutions of semiflexible thermally undulating polymers, like actin, can give rise to a time-dependent shear modulus that scales as $G(\omega) \propto (i\omega)^{3/4}$ . This in turn gives rise to a longitudinal time-dependent displacement proportional to $\Delta t^{3/4}$ in response to a constant applied tension [22]. The latter would naturally lead to $MSD \propto \Delta t^{3/2}$. Therefore a first tempting explanation for this power-law scaling of the peroxisomes' MSD is that elastohydrodynamic relaxation phenomena in the local thermally excited semiflexible polymer environment *within the processes* give rise to an effective time dependent viscosity. However, this seems an unlikely explanation in the S2 cells for several reasons.

First, the peroxisomes moving on the same process, often within 1-2 microns, and in some cases even passing over the same stretch of the process at different times, can exhibit both constant velocity and hyperdiffusive behavior. In some cases, single peroxisomes even switch their behavior from processive linear velocity to hyperdiffusive behavior within the same local environment. This is in sharp contrast to the expectation that spatially close peroxisomes should exhibit a similar environment and therefore similar viscoelastic drag forces. Secondly, taking the almost complete depletion of actin from the processes into account as seen from fluorescent phalloidin staining [7] and the known absence of intermediate filaments from *Drosophila* cells [33] the only long filament to give rise to an elastohydrodynamic response would be the microtubules themselves. However a quick calculation of the maximal thermally stored slack length for the longitudinally aligned microtubules within the processes gives $l_T = L^2 / 6 l_P \approx 5 - 20 nm$ for microtubule length $L \approx 5 - 10 \mu m$ (typical length of the processes) and persistence length $l_P \approx 3 - 5 mm$. Therefore, despite the right scaling behavior, pulling out of the thermally stored microtubule slack length within the processes cannot account for the much larger displacements of several microns observed for the vesicles and the microtubule tips.

Given these observations, a more parsimonious explanation is that non-thermal (motor induced) forces and quenched disorder constraining the microtubule backbones within the cell body generate large backbone undulations. Numerous constraints are imposed by the crowded intracellular environment, forcing the microtubule backbone into an effective highly curved confining tube [32], in particular through entanglement with other microtubules. The large stored length of microtubules (within the cell body) is transmitted over long distances by the virtually incompressible microtubules and projected in the longitudinal direction inside the processes.

The deformations caused by intrinsic or imposed microtubule curvature disorder interestingly give rise to the same longitudinal response $\langle \Delta x \rangle \propto \Delta t^{3/4}$ (i.e. $\langle \Delta x^2 \rangle \propto \Delta t^{3/2}$ ) scaling of the microtubule backbone both for pulling (along its longitudinal direction) and the subsequent free (zero applied tension) relaxation [34]. This indicates that the observed scaling is not an exclusive signature of thermal force—tension competitions, but rather reflects the most generic response of any type of semiflexible filament deformation (intrinsic or imposed from outside) to a tension variation. The fluctuating tensions are induced by multiple molecular motors decorating intracellular cargos and cross-bridging between several microtubules at a time.

The microtubule network actively "animated" in this fashion induces an additional velocity component that adds to the motor driven cargo transport velocities in the microtubule fixed reference frame. In the case of cargos resting with respect to moving microtubules we observe the characteristic 3/2 exponent while in cases of active cargo hauling along (motile or stationary) microtubule the constant velocity motion (quadratic scaling of MSD vs. time lag) eventually dominates over the 3/2 scaling at long times.

The observed predominant 3/2 power-law scaling can be physically understood as the relaxation of many hydrodynamic modes of the microtubule polymer, where a mode with wave number q decays exponentially at the characteristic timescale $t_q \square q^{-4} \eta / B$ [16]. If however only a single wave length L is involved in the relaxation event as in the case of a de-buckling microtubule we expect a purely exponential decay on a single timescale $t \square L^4 \eta / B$ consistent with occasional pure exponential velocity traces as in Fig.1a, blue trace.

Remarkably the characteristic $MSD \propto \Delta t^{3/2}$ -scaling is commonly observed in the motion of many different cargos in several other eukaryotic systems [22]-[25]. However (with the notable exception of the work of Lau et al. [24]) it has been attributed to the local network viscoelasticity *hindering* the vesicle motion in a time dependent manner, rather than to motile microtubules.  Based on two-point microrheology measurements Lau *et al.* [24] suggested that the unusual scaling could be the consequence of a fluctuating background of spatially uncorrelated force-doublets acting throughout the microtubule network.

 As suggested by our data, within the "fluctuating cytoskeleton" picture we can indeed understand the observed back and forth motion as a consequence of a peculiar form of tug of war of many motors competing with each other and with microtubule elastic forces. As opposed to the "local" tug of war of opposite polarity motors on the same vesicle, the "global" tug of war described here allows large numbers of motors distributed along the whole microtubule to exert forces at a time and compete for the direction of microtubule movement. When bent on large scales, the microtubules offer a rather large compliance to the exerted longitudinal and lateral forces, which in turn allows all the motors along their length to act at a time and generate the observed microtubule fluctuations. Switching of motor pulling and microtubule relaxation phases can induce a back and forth motion of the microtubule backbone.

 The documented microtubule motion leads directly to the question of how the cargo motion will be related to it. On short timescales a peroxisome passively adhering to the microtubule would simply follow the microtubule motion. However, on longer timescales (10s of seconds to minutes) the coupling between them might temporarily fail and the peroxisome might unbind from the microtubule. A repetitive binding/unbinding process from the microtubule leads to an eventually diffusive behavior i.e. $MSD \propto \Delta t$ on long enough timescales (longer than the vesicle binding time) [21]. This long time behavior (on timescales > 10 s) is indeed observed for a large portion of motile peroxisomes in the processes (80%)  (cf. Supporting figure) while a smaller portion of them, presumably strongly sticking to the microtubules, shows a confined behavior.  For this mode of motility involving transient binding of cargos to moving microtubules which eventually leads to a long-range dispersion, we suggest the term

"hitchhiking". Exploiting this simple mechanism, even cargos devoid of active motors can be efficiently dispersed throughout the entire cell [21].

In light of the presented data, a simple model of bidirectional transport on stationary microtubules does not adequately describe organelle translocation in *Drosophila* S2 cells. We demonstrate that besides being tracks for motors that directly haul cargos, microtubules can transmit the force of distant motors onto a cargo over large separations. This implies a mechanical non-locality of the cytoskeleton since a longitudinal pulling strain in an almost stretched microtubule is essentially instantaneously transmitted over long distances. Furthermore, microtubule motion on intermediate timescales (tens of milliseconds to several seconds) can be understood as a consequence of pulling-out the slack length of microtubules induced by random constraints and motor forces along its entire length.

Presently it is an open question to what extent microtubule movement contributes to the phenomenology of bidirectional organelle transport in other cellular systems besides the processes of drosophila S2 cells we employed. However, it remains an attractive hypothesis that this mechanism might be a commonly employed in other eukaryotic cell types. This question as well as the precise molecular mechanisms that drive microtubule movements in the cytoplasm is the subject of our future investigations.

## *Materials and Methods*

***Drosophila* cell culture and stable cell line selection.** *Drosophila* S2 cells were maintained in Schneider medium supplemented with 5% Fetal Bovine Serum, 0.1 mg/ml Penicillin and 100 U/ml streptomycin at $25^0C$ in a humidified incubator. To select a stable cell line co-expressing EGFP-SKL (peroxisomal marker), and mCherry-tubulin, S2 cells were co-transfected with pAC-EGFP-SKL (a gift of Gohta Goshila, Nagoya University), pMT-mCherry-α-tubulin and pCoHygro (Invitrogen) in 20:20:1 molar ratio using Cellfectin (Invitrogen). 300 μg/ml of Hygromycin was added to normal growth medium 48 hours after transfection. The expression of tagged proteins was confirmed by fluorescence microscopy after 8-hr induction with 0.1 mM copper sulphate. Cells for microscopy were plated on Concanavalin A-coated coverslips in the medium containing 5 μM Cytochalasin D to depolymerize actin as described in [12]

**Imaging.** Two-color imaging of peroxisomes and microtubules was performed using a 100 X 1.49NA lens and 1.5 X intermediate magnifier on a Nikon U-2000 inverted microscope equipped with a Perfect Focus system (Nikon Instruments, Melville, NY) and Cascade II EMCCD (Ropper Scientific) driven by Metamorph software.  A 100 W halogen light source was used for fluorescence excitation to minimize photobleaching and phototoxicity. Fast TIRF single-color imaging was performed as described by Kural *et al.* [12].

**Vesicle tracking and trajectory analysis**. Vesicle tracking was performed with a custom Gaussian centroid fitting algorithm as described by Kural *et al.* [12] The trajectories of EGFP labeled peroxisomes inside S2 cell processes were rotated and their dominant components along  the process direction were analyzed. Often (in ~40%) peroxisome trajectories inside processes exhibited a localized motion with no clear axis of motion indicating a rigid attachment to resting microtubules or other structures. To determine the mean-square displacement exponent, we focused on the motile fraction of vesicles with large aspect ratio trajectories. The motile fraction was defined by following two criteria: a) The aspect ratio of the longest and the shortest axis of the peroxisome trajectory over a time period of 5 s was larger than 3, and b) The total absolute peroxisome displacement over 5s was larger than 100nm.

 Only long trajectories (>5000 frames, 5 sec) with low non-specific white noise levels (MSD exponent over first 30 ms larger than 0.1) were included in the analysis (N=36).

**Microtubule tracking and relative sliding analysis**. Microtubule contours were tracked with a semiautomatic ImageJ plugin NeuronJ [9] and the arc lengths of the digitized trajectories were calculated and analyzed by a custom Matlab routine. The relative sliding speeds of microtubules with respect to each other were evaluated by analyzing the rate of change of the arc-length difference between two neighboring positions at which microtubules converged together.

**Acknowledgements**.  This work was supported in part by NIH grant GM-52111 from NIGMS to V.I.G., AR44420 to P.R.S., and NSF Grants DMR04-04674 and DMR04–25780 to P.C.N, and an NSERC of Canada grant to A.E.X.B. I.M.K was supported in part by the Max Planck Society.

# Figures

**Figure 1. Unusual dynamical features of perxisomes moving in S2 processes.** (a) Typical displacement vs time of peroxisomes along the microtubule direction, characterized by non-constant velocities, indicates the involvement of microtubule elastohydrodynamics. In some cases the initial relaxations were extremely rapid (~12 $\mu$m/s) but quickly slowed down and were well fitted by single exponentials (blue curve and the inset, 610 nm total length release, 60ms relaxation time). (b) Two peroxisomes (inset, left) move almost perfectly in concert over a large time window (cf. supporting movie 1). The vertical axis is displacement of the peroxisome center along the axis shown as a dashed line in the left inset. The initial displacements are shifted to facilitate the comparison. Inset (right): the velocity cross-correlation (y-axis) of the two particles as a function of the coarse-graining (CG) time (the time interval over which the mean velocities are evaluated), horizontal axis. On short times (<30ms) the vesicles undergo individual uncorrelated dynamics whereas on longer times (>100 ms) they become strongly correlated. (c) The mean-square-displacement (MSD) of the vesicles from b) vs. time shows power-law scaling with a subdiffusive exponent (0.82 and 0.70) for times < 30 ms and from 100 ms to 3 s a hyperdiffusive exponent (1.47 and 1.49). Inset: MSD vs. time of representative traces, (Thick lines indicate the slopes 0.5 and 1.5) (d) The distribution of short-time and long-time scaling exponents of the MSD from N=36 peroxisomes ($\alpha$=0.59 +- 0.28 and $\beta$= 1.62 +- 0.29) .

**Figure 2. Microtubules undergo massive rearrangement.** Left panel: Two color fluorescence image of a Drosophila S2 cell with microtubules shown in red and peroxisomes in green (see supplemental movie 2.mov, Scale bar = 10 µm). Right panel: frames from movie S1 with frame number indicated at the bottom (frame rate 1 s-1). (a) A microtubule is seen buckling out of a bundle (arrowhead) near a peroxisome. (b) More extreme buckling is also observed in some processes (buckling microtubule is traced in white). In this case, the buckling microtubule's relative slack changes at an average rate of 1.0µm s-1. (c) A microtubule loop shown with an arrow head is transported down the length of a process at 0.8 µm/s. Given microtubules' inextensibility, all of these events require significant microtubule sliding within bundles.

**Figure 3. Peroxisomes simultaneously bind multiple microtubules or move with them in concert.** (a) A peroxisome (indicated by the arrowhead) dynamically clamps two microtubules as it moves along both, releases one of them in frame 4, and continues to the right in frame 7 (passing another immobile peroxisome) while the microtubules splay apart. (Scale bar = 2.5 µm. Frame rate = 1 s$^{-1}$). (b) A peroxisome coincides with a strongly dynamically rearranging microtubule bundle kink. The bundle splits into several microtubules which converge at the vesicle position (frame 31, arrowhead). Scale bar = 5 µm. Frame rate= 5.7 s$^{-1}$. (c) Kymograph of the peroxisome and microtubule shown on the left. The microtubule tip and the peroxisome move together (correlation coefficient =0.92).

**Figure 4. Microtubules form "hubs" that can persist for minutes.** Left panel: Drosophila S2 cell. Scale bar represents 10 µm (see supplemental movie 4.mov). Right panel: closer view of hubs (indicated by arrowheads). Frame number is indicated in white. Frame rate is 1/s. While the associated microtubules move over microns and change shape the crossing points ("hubs") remain stably associated (pair-wise crossover points remain confined to each other to within <500 nm) suggesting a binding mechanism.  In part (b) a microtubule unbinds from a hub (between frames 1 and 36) and another microtubule binds to the same location and stays associated (between 36 and 71).

**Figure 5. Model of S2 cell cytoskeletal fluctuations.**  Characteristic events and their microscopic interpretation, a+b): Multiple motors of different polarity operating on different vesicles and multiple microtubules lead to single microtubules buckling away from bundles or bundles globally deforming on the substrate (cf. Figures 2 a-b and 3 b) . c) Cargos carrying multiple motors dynamically crosslink multiple microtubules ("hubs") and induce lateral and longitudinal fluctuating forces (cf. Figures 3a, 4a-c). d) Multiple vesicles bind statically to a single microtubule and experience passive "hitchhiking" along the longitudinally fluctuating microtubule (cf. Figures 1b, 3c).

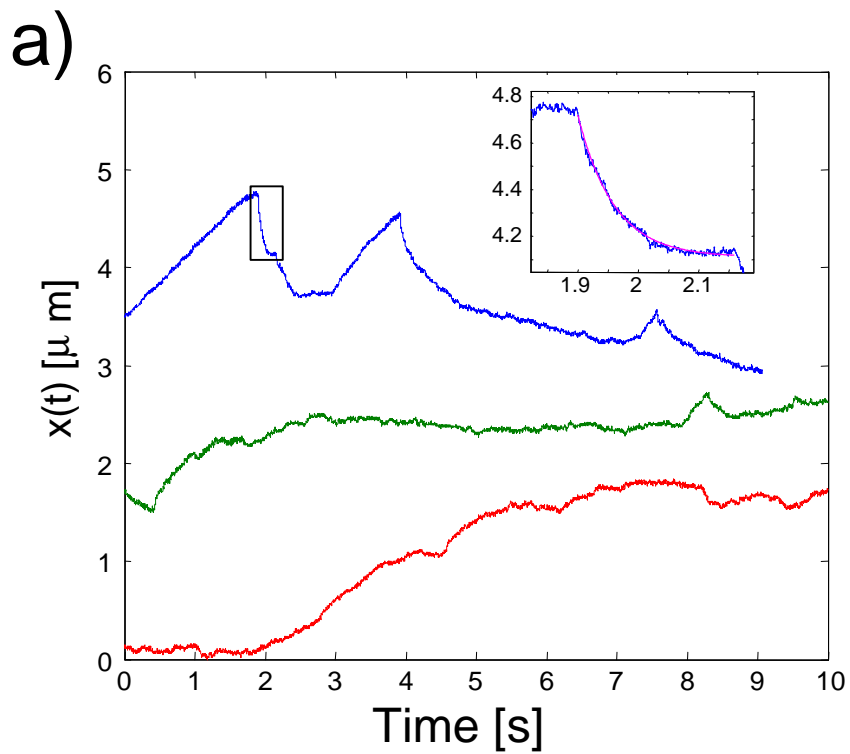

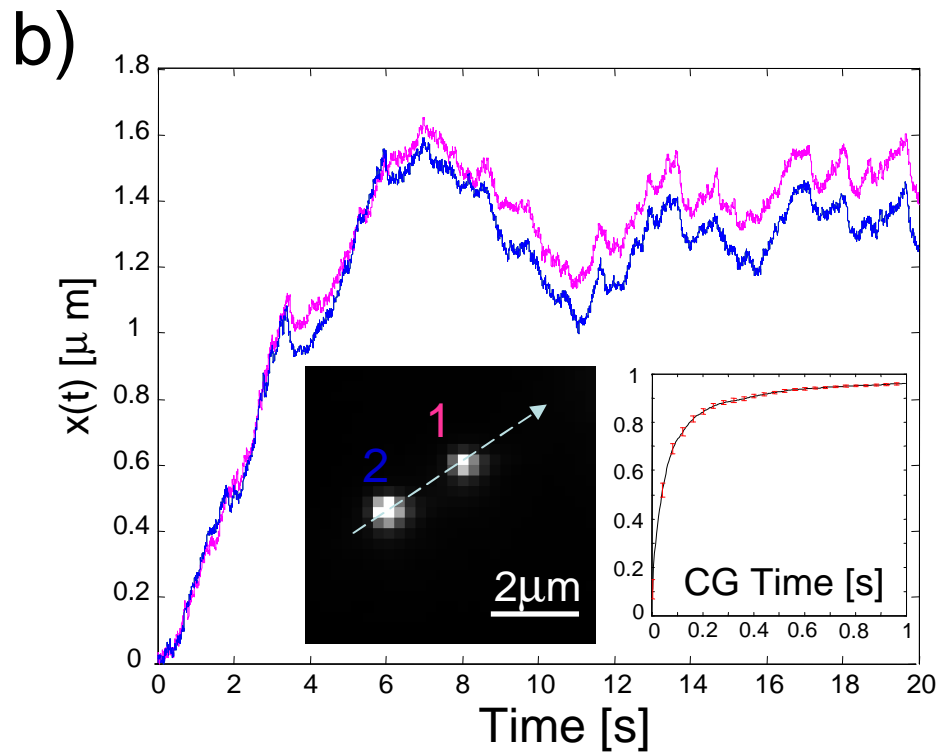

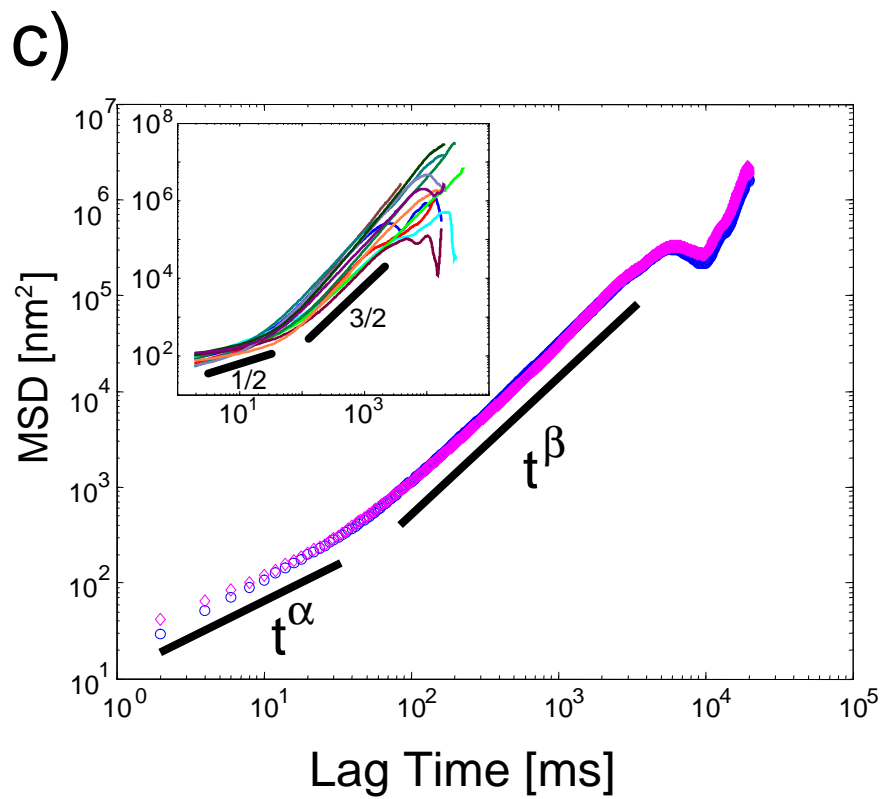

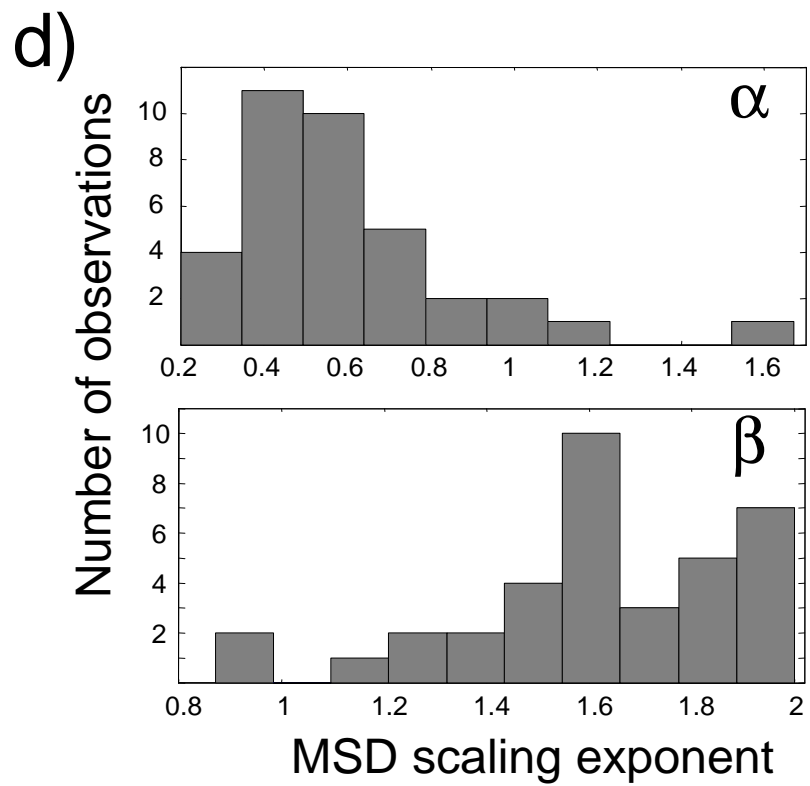

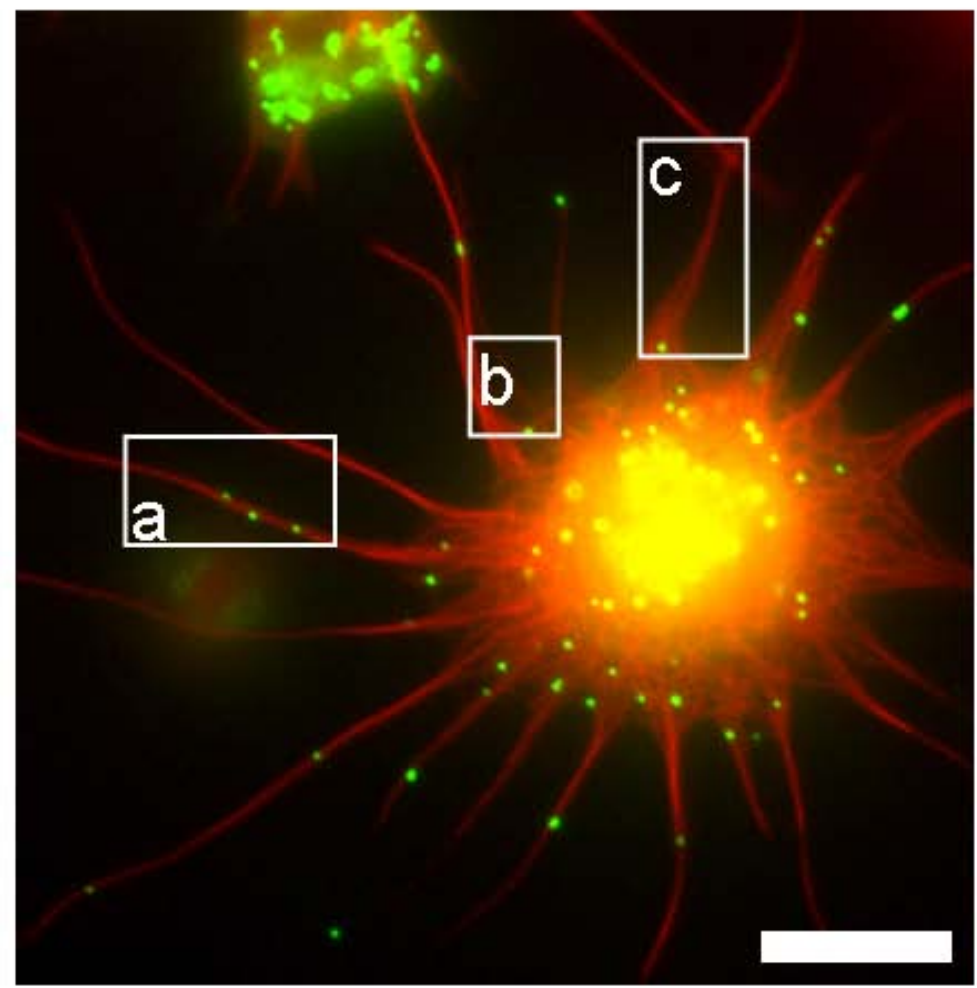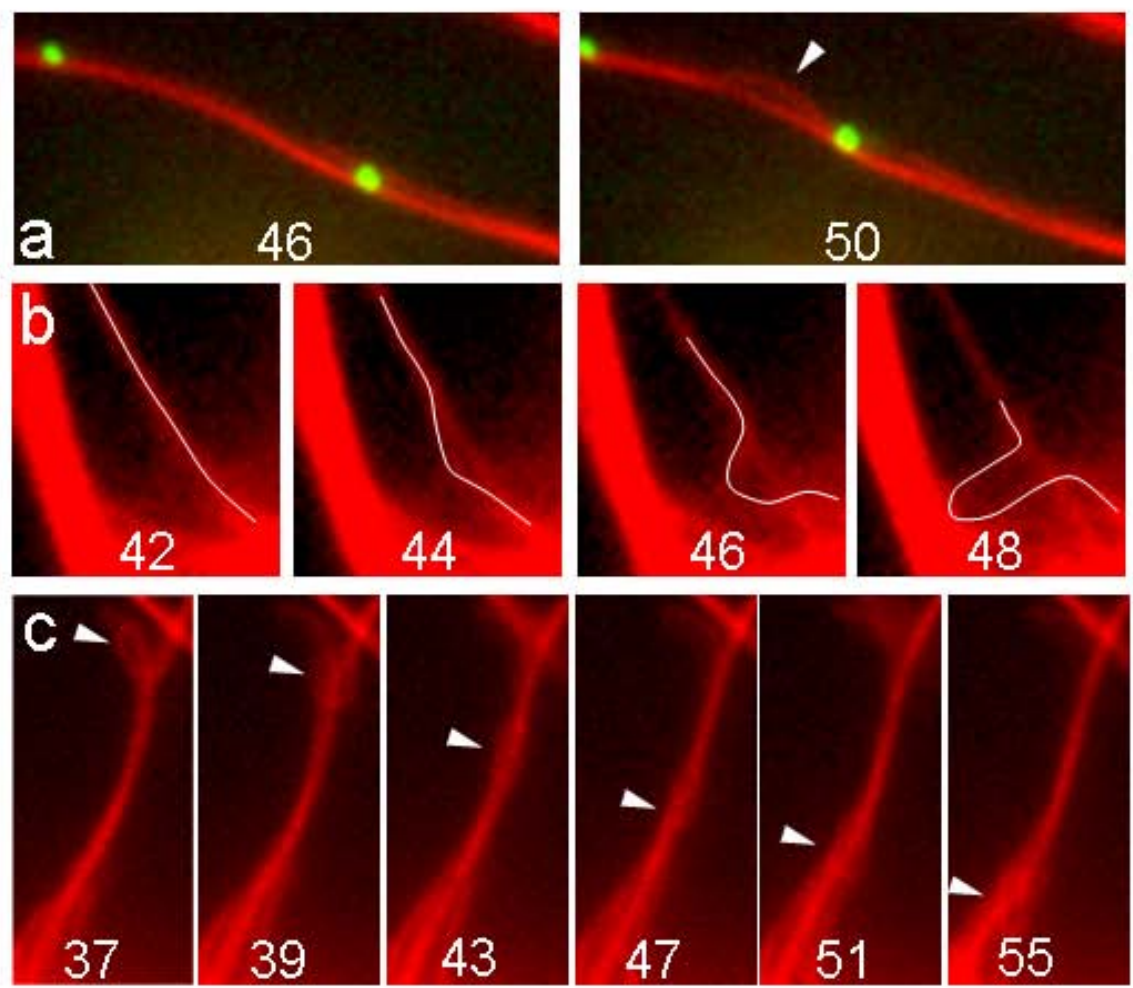

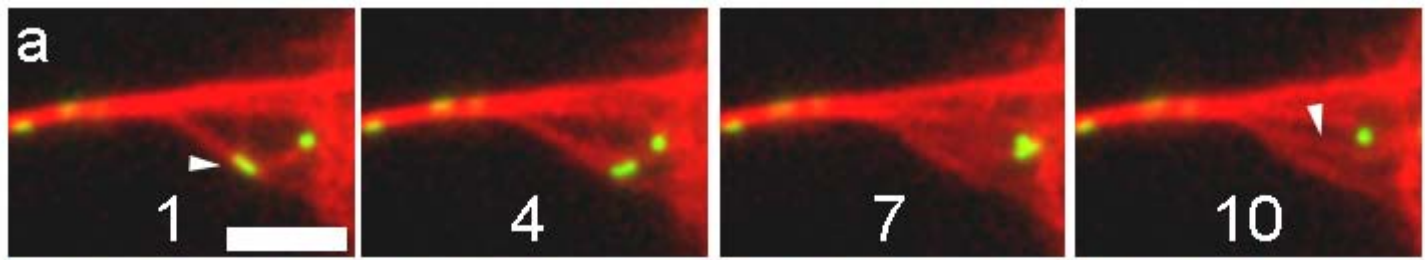

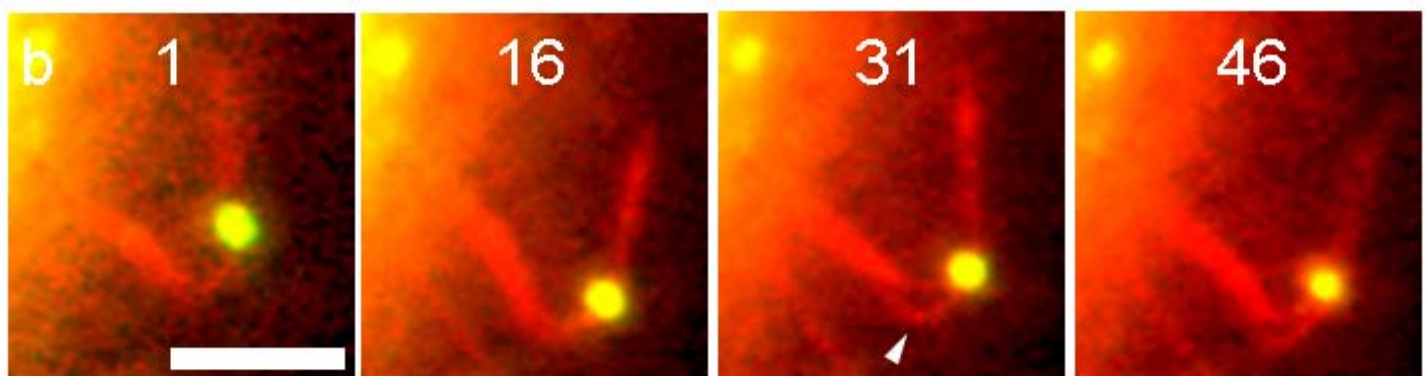

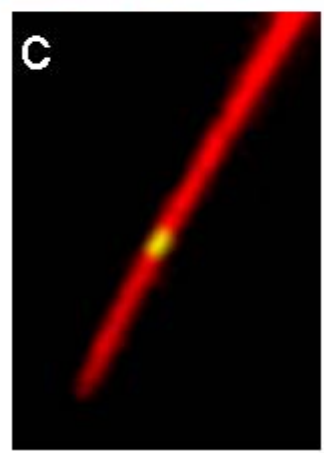
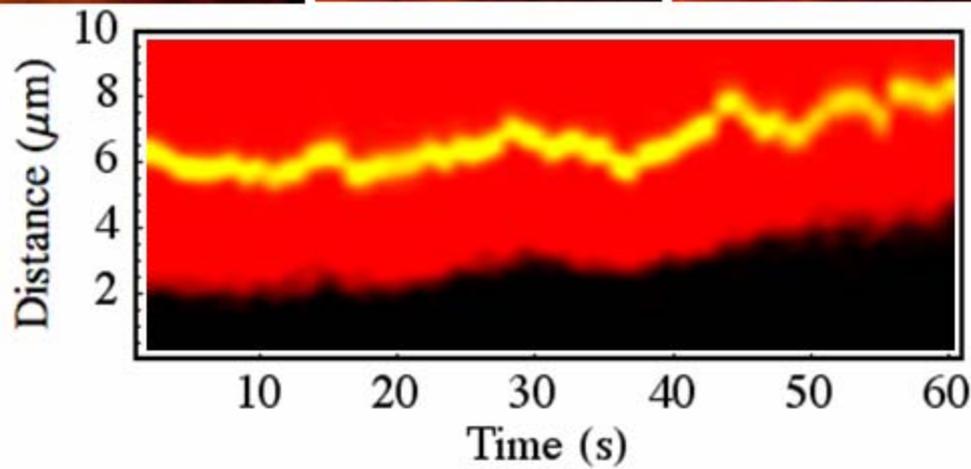

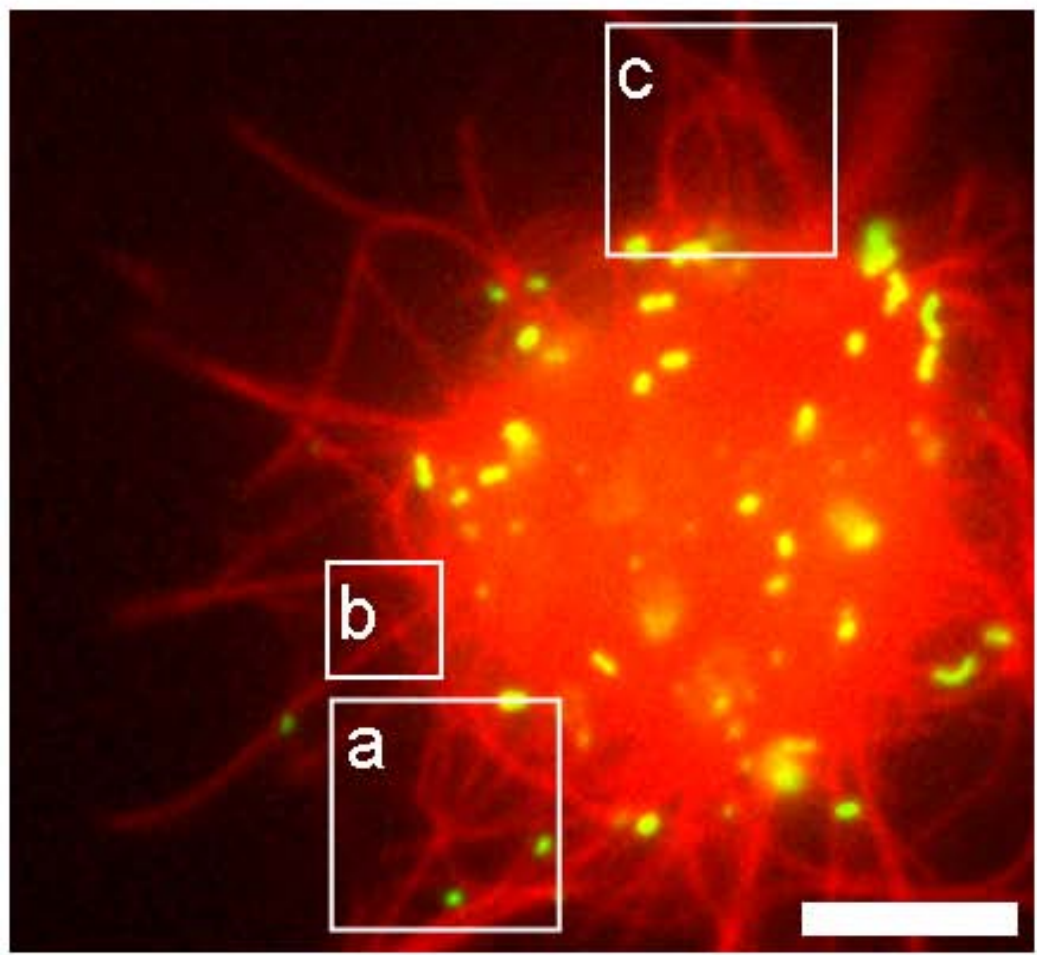
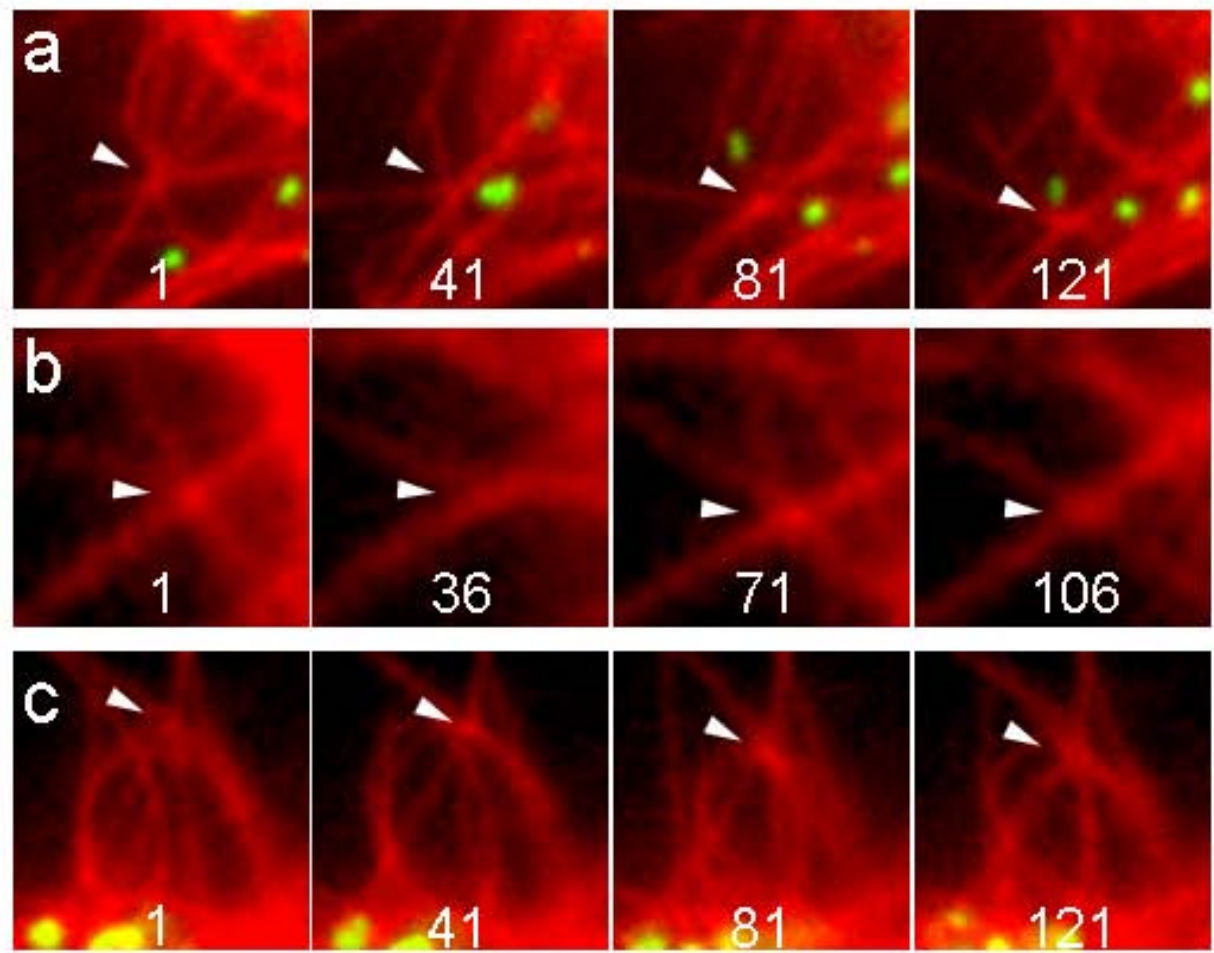

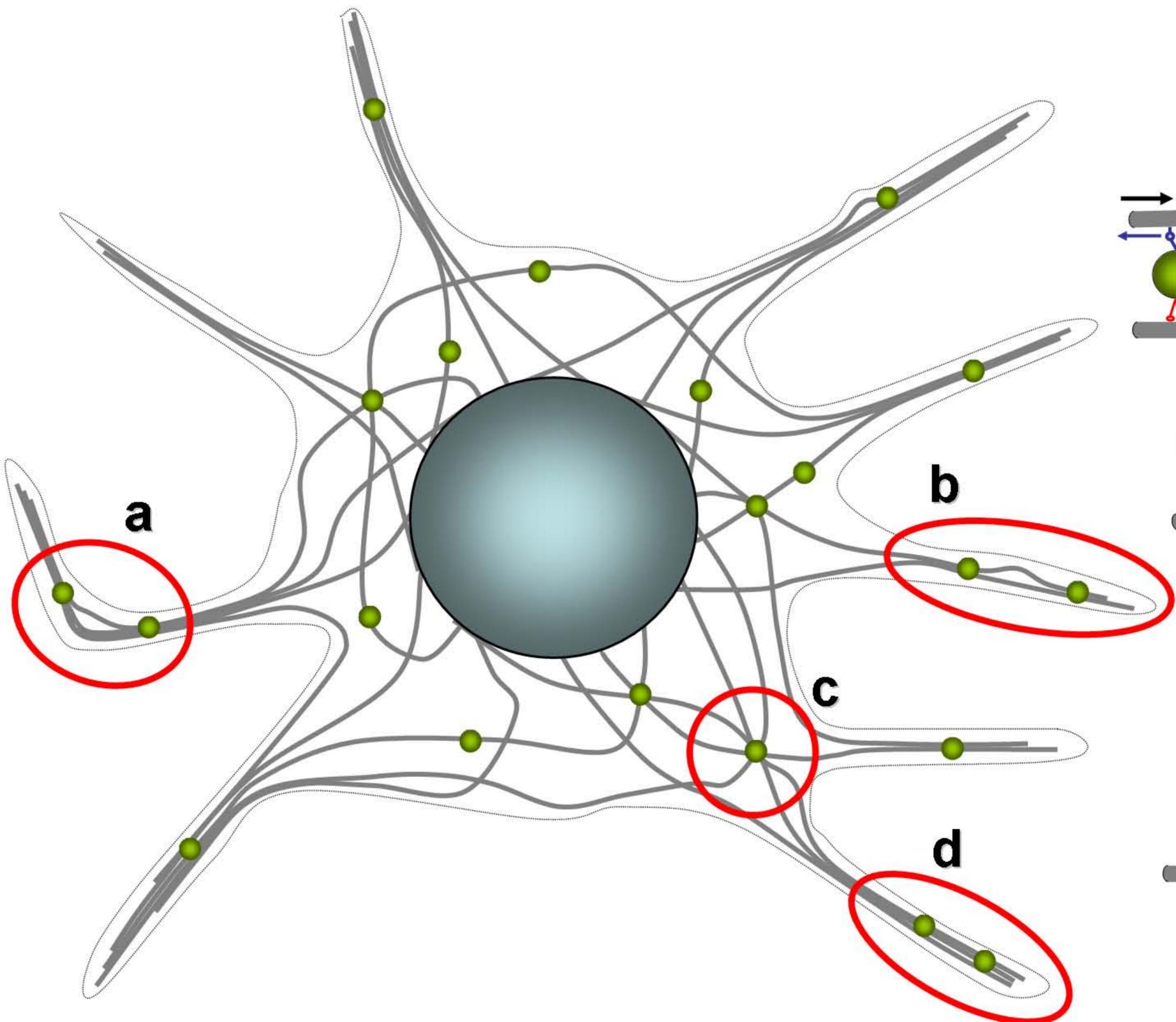

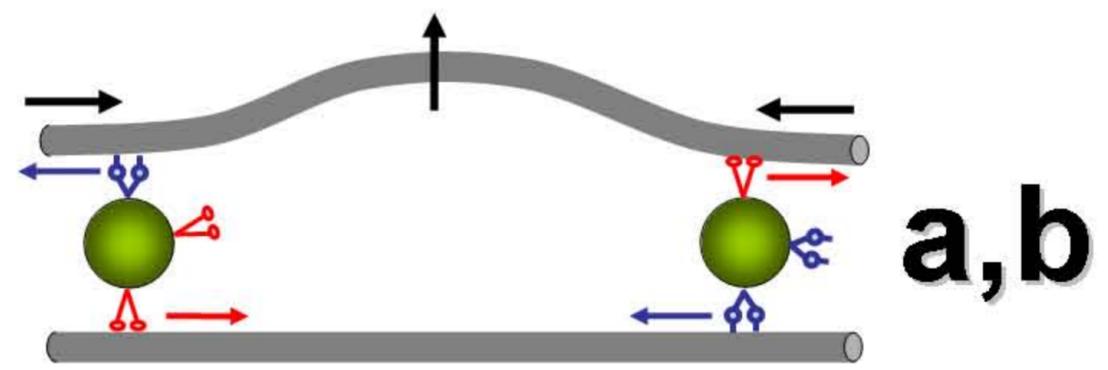

**a,b**

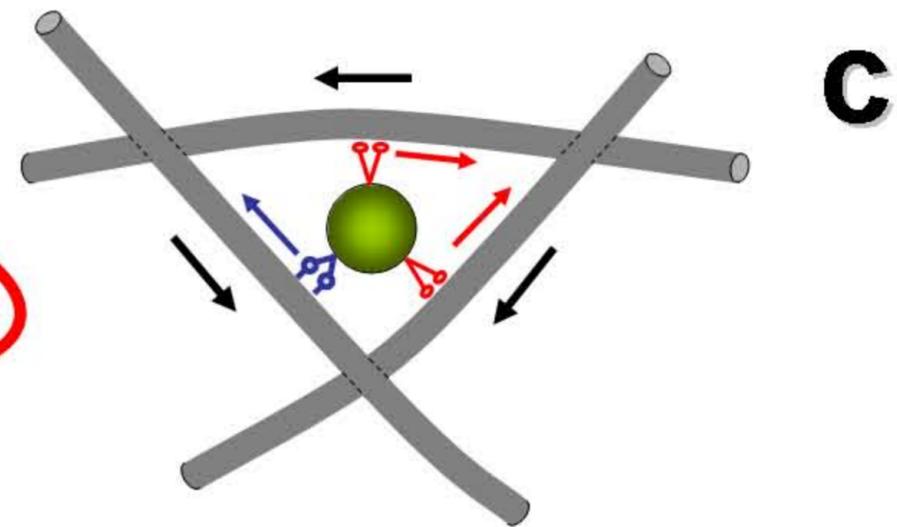

**c**

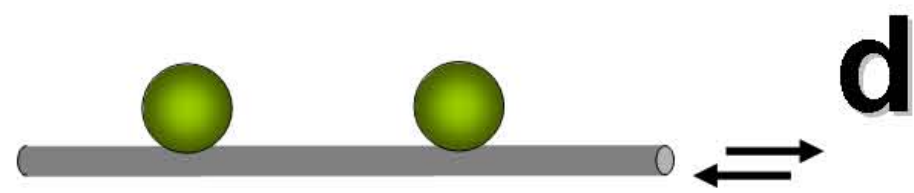

**d**